\documentclass[usenatbib,usegraphicx]{mn2e}
\usepackage{amssymb}

\title[Spectroscopy of the globular cluster black hole in NGC 4472]{Multi-epoch spectroscopy of the globular cluster black hole in NGC 4472}
\author[I C. Shih et al.]{I.C. Shih$^{1}$\thanks{E-mail:shihs@pa.msu.edu}, T. J. Maccarone$^{2}$, A. Kundu$^{1}$, and S. E. Zepf$^{1}$ \\
$^1$Department of Physics \& Astronomy, Michigan State University, East Lansing, MI 48824, USA \\
$^2$School of Physics \& Astronomy, University of Southampton, Southampton, SO17 1BJ}

\newcommand{\aap}{A\&A}
\newcommand{\apj}{ApJ}
\newcommand{\apjl}{ApJL}
\newcommand{\apjs}{ApJS}
\newcommand{\araa}{ARA\&A}
\newcommand{\mnras}{MNRAS}
\newcommand{\nat}{Nature}
\newcommand{\pasj}{PASJ}

\begin{document}

\maketitle
\begin{abstract}
We present a study of the X-ray spectral properties of the highly variable X-ray emitting black hole in a globular cluster in the elliptical galaxy NGC~4472. The XMM-Newton spectrum of the source in its bright epoch is well described by a multiple blackbody model with a characteristic temperature $kT_{in}\approx$ 0.2 keV. The spectrum of an archival Chandra observation of the source obtained 3.5 years before the XMM data gives similar estimates for the blackbody parameters. We confirm that the fainter interval of the XMM-Newton observation has a spectrum that is consistent with the brighter epoch, except for an additional level of foreground absorption. We also consider other possible mechanisms for the variability. Based on the timescale of the X-ray flux decline and the estimated size of the X-ray emission region we argue that an eclipsing companion is highly unlikely. We find the most likely means of producing the absorption changes on the observed timescale is through partial obscuration by a precessing warped accretion disk.

%We conclude that our initial suggestion of the precession of a warped accretion disc which partially absorbs the central X-ray emission region is the most likely mechanism for producing the temporal variation in foreground absorption.
\end{abstract}


\section{Introduction}
\citet[hereafter MKZR07]{Maccarone:2007lr} identified a luminous
(4$\times$10$^{39}$ erg s$^{-1}$) X-ray point source, XMMU
122939.7+075333, in a globular cluster around the Virgo elliptical
galaxy NGC~4472. Its count rate varied by a factor of seven over a few
hours during a 28 hour observation made by the X-ray Multiple
Mirror-Newton (XMM-Newton) satellite. This X-ray luminosity is well
beyond the Eddington limit ($\approx10^{38}$ erg s$^{-1}$) for
accretion onto a neutron star.  While accretion powered pulsars are
sometimes seen to exceed the Eddington limit for long stretches, low
magnetic field neutron stars can tolerate only very short excursions
above the Eddington luminosity. High magnetic field neutron stars of the former variety are not expected to exist in the old stellar environment of
a globular cluster.   The strong X-ray variability of this source rules out the
possibility that the X-ray emission could come from multiple sources. In fact variability is one of the only ways to show convincingly
that an extragalactic globular cluster X-ray source is a black hole
\citep[e.g.][]{Kalogera:2004fk}.  Finally, the optical counterpart is
spectroscopically confirmed to be a globular cluster with a
redshift appropriate for a cluster in NGC~4472, eliminating the
possibility that this source might be a background AGN \citep[for details of the optical spectroscopy, see][]{Zepf:2007nr}.

This object exhibits two remarkable features - its X-ray intensity
shows very high amplitude variability, and its X-ray spectrum peaks at
a lower X-ray energy than the spectra of typical X-ray sources in
galaxies. The initial spectral analysis indicated that the variation
is predominantly at lower energies ($<$ 0.7 keV) and hence it may be
due to the increase of the absorption column in front of the X-ray
source (MKZR07). While the softer X-ray spectrum might suggest an accreting
intermediate mass black hole (IMBH) in the object \citep{Miller:2003}, as noted in MKZR07,
other mechanisms are also capable of producing a soft X-ray spectrum
at high luminosity from stellar-mass black holes e.g. accretion
onto slim discs at very high accretion rates \citep{Abramowicz:1988fk,
Begelman:2006kx}. Similar phenomenology has been seen in other systems in the past. For example, the Galactic black hole candidate V404 Cyg showed rapid variations in its foreground absorption level \citep{Oosterbroek:1997uq}, while cool accretion disks have been found in a subset of ultraluminous X-ray sources \citep[see][]{Stobbart:2006}. 

%On the other hand, the object's luminosity and soft X-ray spectrum resembles the main characteristics of Ultraluminous X-ray Sources , which also exhibit variation in X-ray luminosity, although the physical mechanism for the variation in individual sources may be different, i.e. ULXs in NGC 1313 \citep{Mizuno:2007fk}.

In this paper we use the XMM-Newton data in conjunction with archival
ROSAT and Chandra observations to study further the X-ray spectrum and
variability of XMMU 122939.7+075333. We present the spectral analysis
results in section 3, extending the spectral analysis already presented
in MKZR07. We show that the luminosity and spectrum observed in the Chandra data is similar to the the bright interval of the XMM-Newton data, and confirm that the X-ray flux decline is most likely due to an increase in the absorption column. In section 4, we use the observed decline timescale of
$\sim$ 13 ks from the bright epoch to the faint epoch of the
XMM-Newton observation to put constraints on the geometry of the foreground absorption.

\section{Observations and data reduction}
\subsection{XMM-Newton}

NGC 4472 was observed for 1.1$\times$10$^{5}$ seconds by the
XMM-Newton on 1st January 2004. All three European Photon
Imaging cameras (EPIC), including two Metal Oxide Semi-conductor (MOS)
and pn, were simultaneously operating with full-frame mode and thin
filter.  The data analysis procedures are essentially the same as
those used in MKZR07, but we outline the analysis
procedure in the interest of completeness. We use standard procedures
to analyze the XMM-Newton data. The Observation Date Files (ODF) were
reprocessed using the latest version of Science Analysis Software
(SAS) v.7.0 and Current Calibration Files (CCF) to create an updated
calibrated and concatenated event lists. To generate science products
such as images, lightcurves and spectra, a set of selection criteria
were applied to the concatenated event files \citep{Loiseau:2006}. High background flares interval is filtered out if the count rate at the pattern 0 is higher than 0.35 cts/s and 1.0 cts/s in the MOS and the pn cameras, respectively. As
higher number patterns are not created by X-rays or are highly
contaminated by cosmic rays, only event patterns 1-12 and 0-4 are
selected for the MOS and the pn camera, respectively. FLAG==0 omits
parts of the detector area like border pixels, columns with higher
offsets, etc., and \#XMMEA\_EP, \#XMMEA\_EM are used to filter out
artifact events. The good time intervals (GTIs) of the bright and faint parts of the data is defined by 0.015 cts/s. For spectral analysis, the created spectra were regrouped into bin of at least 20 photons.  

\subsection{Chandra} 
NGC 4472 was observed by Chandra ACIS-S on 12 June 2000 for
$\sim$40,000 seconds. The data were obtained from the Chandra Data
Archive and reprocessed to generate improved calibrated level=2
event files using the Chandra Interactive Analysis of Observations
(CIAO) software package v.3.3 \citep{Fruscione:2006}.

The globular cluster black hole source is located on the ACIS-S2 (CCD 6). After the time intervals with background flares were
filtered, we obtained a total exposure time of $\sim$38 ks. An
examination of the lightcurve shows that the X-ray intensity of the
object is steadily bright. The CIAO {\tt dmextract} task is used to
create source and background spectra, a circular background region is
selected in the same CCD where no significant X-ray source is
presented. The {\tt mkacisrmf} and {\tt mkarf} tasks create the
instrument models of RMF and ARF, respectively. Finally, in order to
have good statistics, the source and background spectra are regrouped
to have a minimum number of 15 and 20 counts per new channel,
respectively.

\subsection{ROSAT} 
ROSAT observed NGC~4472 for about 27,000 seconds of live time in June
and July of 1994. The quality of spectra from the ROSAT High
Resolution Imager is quite poor, so these data cannot be used for
spectral fitting to test models.  XMMU 122939.7+075333 was 
already reported as an intermediate luminosity X-ray object
\citep{Colbert:2002uq} -- it is IXO 60 in their catalog. They
estimated a source luminosity of $10^{39.9}$ ergs~s$^{-1}$ from 2-10
keV, based on an assumed spectrum of a power law with photon index
$\Gamma=1.7$, where $\frac{dN}{dE}$$\propto$$E^{-\Gamma}$ -- a
considerably harder spectrum than what we have fit from the XMM-Newton
data during the bright epoch.  The source count rate of about
$4\times10^{-3}$ cts/sec is consistent within a factor of a few of
both the fainter and brighter states seen from XMMU 122939.7+075333
with XMM -- because a harder spectrum was assumed by \citet{Colbert:2002uq} than what we have found, they infer a higher luminosity for the source
than what we have found despite the fact that the XMM-Newton spectrum,
folded through PIMMS for ROSAT, gives a higher count rate than what
the ROSAT HRI saw.  The ROSAT data are useful for helping to
demonstrate that the source has likely been on as a bright source for
at least the 9 year period between the ROSAT and XMM observations, but
are not sufficient for helping to determine whether the bright or
faint part of the XMM-Newton observation is typical of the source's
behaviour.

\section{Spectral analysis}
The primary goals of the analysis are to investigate the nature of the
flux decline in the XMM-Newton observation, and to compare the XMM and
Chandra spectra. In MKZR07, a restricted set of spectral fits were
presented, and fits were presented only to the XMM-Newton data. It
was shown there that the spectrum extracted from the bright part of
the XMM-Newton light curve is consistent with the multiple
blackbody disc model (often referred to as a multi-color disc or MCD) which models a multiple blackbody from an accretion disc,
parameterised in term of the temperature (keV) at inner disc radius
\citep{Mitsuda:1984fk}, with
foreground Galactic absorption only. The spectrum during the
faint part of the observation was found to be consistent with the same underlying
model, but with a higher value for the absorption.  Here we present a
more detailed discussion, considering alternative models for the
spectrum, and also fit the Chandra spectrum. The effect of Galactic absorption was taken into account by using {\tt phabs} in {\tt XSPEC} \citep{Arnaud:1996aa}. The Galactic absorption at the direction of the
object is $N_{H}$ is $\sim$1.6$\times$10$^{20}$cm$^{-2}$ which was
frozen in most of the fits. 

First the spectra of the Chandra data and the XMM observations in both the bright and faint epoch were fitted with an absorbed MCD model. The fitting result is marginally acceptable for
the XMM-Newton spectra but is statistically a poor fit to the
Chandra spectrum. Two notable features, however, can be drawn from the
result.  First, the peak temperature ($kT_{in}\approx$0.2 keV) of all
the spectra is significantly lower than typical X-ray binaries
($kT_{in}\approx$1.0 keV).  Second, there appears to be a ``hard
tail'' in the $>$ 1 keV band, indicating that an additional second
component may be required.

Next we fit the X-ray spectra (0.1-10 keV) with a MCD plus power-law model, as is often done for low mass X-ray binaries. It shows a great improvement for the Chandra spectrum and a slight improvement for the bright epoch XMM-Newton one. However it is worth mentioning that the Chandra data are taken far off axis, where the response matrix is not well calibrated, so that the difference should be taken cautiously{\bf \footnote{see Chandra calibration http://cxc.harvard.edu/cal/}}. To verify that the power-law component really is required for the bright epoch XMM-Newton spectrum, we performed the f-test by comparing the values of $\chi^{2}$ and numbers of degrees of freedom (d.o.f.) from both models. The null hypothesis probability for adding the power law component is $\approx$ 10\%, indicating that the evidence for the extra component is suggestive but not conclusive. For the faint epoch XMM-Newton data, the parameters of the additional component, i.e. $\Gamma$, are poorly defined because of very large error contours ($<$68$\%$ confidence level), and it is most likely due to few data points in the $>$ 1 keV band.

The radial temperature dependence of the MCD model follows a power-law form 
$kT\propto$R$^{-\rho}$, where $\rho$=0.75. The peak temperature refers
to the temperature at the inner accretion disc radius. The
characteristic temperature derived from the ULXs' spectra has led some
to suggest that an intermediate mass black hole (IMXB) is contained in these
sources \citep{Miller:2003}, since more massive black holes will have
lower inner disc temperatures at a given luminosity, under the
assumptions that go into the MCD model \citep{Shakura:1973yj}. On
the other hand, an $\rho$-free multiple blackbody disc
model, assuming super-Eddignton accretion on to a stellar-mass black
hole (i.e. slim disc), can also explain the same spectra
successfully \citep[i.e.][]{Vierdayanti:2006}. When the source enters the
``slim disc'' or optically thick advection-dominated accretion disc state,
$\rho$ reduces to $\approx$0.5. Thus we tried to fit the spectra with
{\tt DISKPBB} model in {\tt Xspec}. For the bright epoch XMM-Newton and for the Chandra data, the parameter values for this model converge to values quite similar to those obtained from the MCD model. For the faint epoch XMM-Newton data, the model produces a good fit without variation of the absorption, but the value of $\rho$ required is 1.0. This is indicative of an accretion disk with a steeper temperature gradient than in a standard Shakura-Sunyaev accretion disk, a result with no clear physical interpretation.

We then consider that the X-ray flux decline may be due to an increase
of absorption column $N_{H}$ as suggested in MKZR07. An absorbed MCD
model with frozen $kT_{in}$ and normalization derived from the bright epoch
XMM-Newton spectrum was fitted to the faint epoch XMM-Newton one. The
result clearly shows that the absorption column increases by an order
of magnitude from bright to faint epochs, and that the fit is
acceptable, despite having only one free parameter, since the other
parameters were specified by the fit to the bright epoch. 

Conclusively, the absorbed MCD model is our best effort to describe the XMM-Newton spectra in the $<$ 1 keV band. For bright epoch XMM-Newton, the f-test suggests that a second component might be necessary to the spectrum in the $> $ 1 keV band, but the relatively low single-to-noise data in this energy band prevent further constraining of the parameters. 

\begin{figure*}
\begin{center}
\includegraphics[width=170mm]{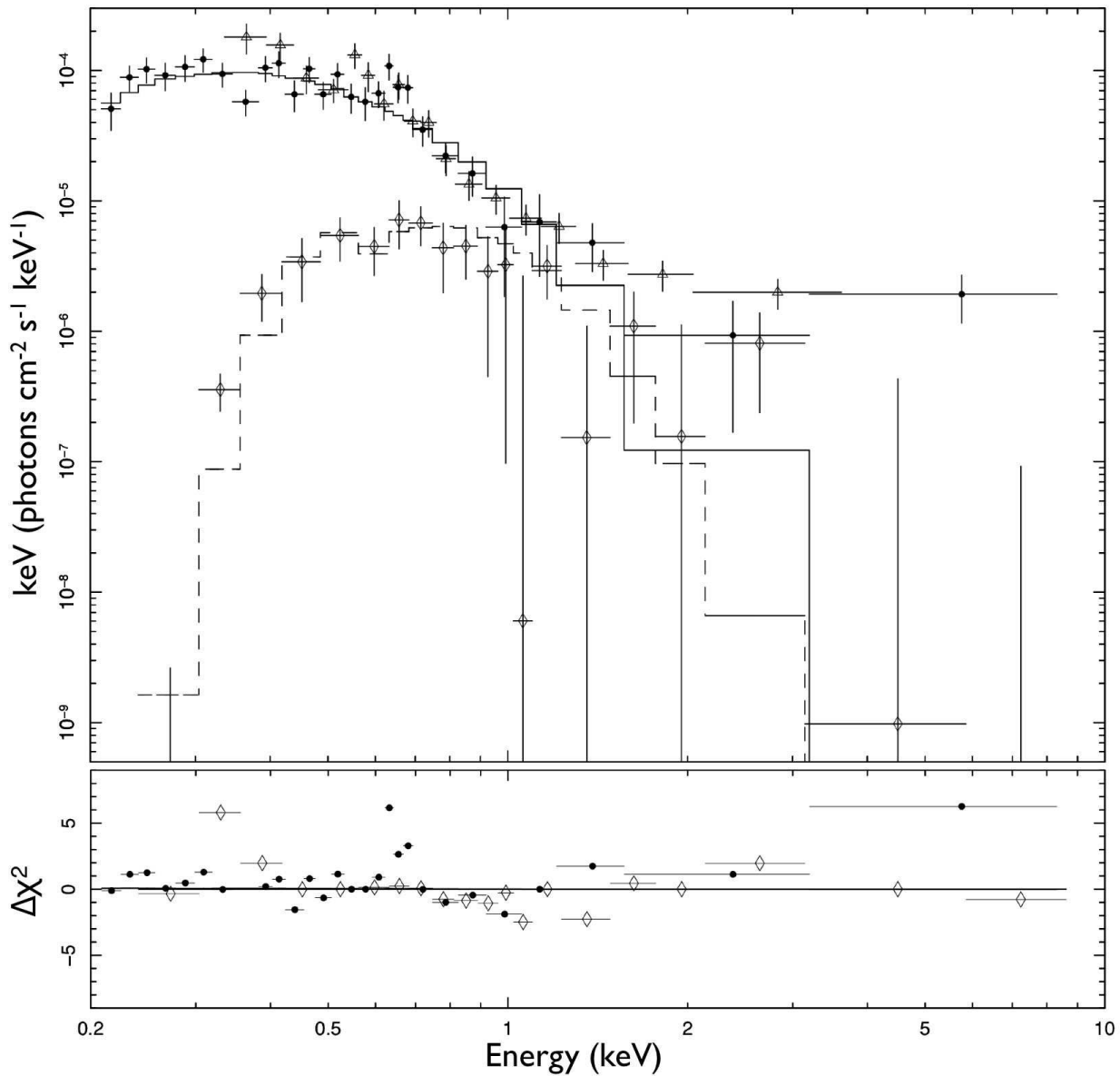}
\caption{XMM-Newton EPIC pn spectra and $\Delta \chi^{2}$ residuals for an absorbed MCD model fit for bright (solid circle, solid line) and faint (open diamond circle, dashed line) epochs. The Chandra spectrum (open triangle) is also overlapped to demonstrate its likeness to the bright epoch XMM-Newton data.}
\label{XMM-Newton spectra and fit}
\end{center}
\end{figure*}

Table \ref{fixed absorption number} lists details of spectral fitting results discussed above, and Figure \ref{XMM-Newton spectra and fit} plots two epochs of the XMM-Newton spectra and the best fit of the single absorbed MCD model.

%\begin{figure}
%\begin{center}
%\includegraphics[width=84mm]{cxc_bh.eps}
%\caption{Chandra ACIS-S count rate spectrum and $\Delta \chi^{2}$ residual for an absorbed MCD+power-law model fit for the black hole candidate.}
%\label{spectrum A fitted by two component model}
%\end{center}
%\end{figure}

\begin{table}
\caption{Spectral fitting results. Numbers in parenthesis emphasize that the parameters are frozen during the fitting, and all errors quoted in the table are bove 90$\%$ confidence level. The error contours may be slightly underestimated for the fits where $\chi^{2}_{\nu}$ is $\sim$1.5. All the fits have a null hypothesis probability of more than 1$\%$ except the Chandra fit, for which there are likely to be significant systematic errors unaccounted for, since the spectrum is taken far off axis. If $\chi^{2}_{\nu}>$ 2, then {\tt Xspec} does not provide error estimation. }
\label{fixed absorption number}
\begin{tabular}{@{}lcccc} 
\hline\hline
Parameter & XMM & Chandra & XMM$^{1}$ & XMM$^{2}$ \\
 & bright &                & faint & faint \\
\hline\hline\\
DISKBB$^{3}$ & & & &\\
$N_{H}$$^{5}$ & (0.016) & (0.016) & (0.016) & 0.35$\pm^{0.05}_{0.04}$ \\
$kT_{in}$$^{6}$ & 0.16$\pm$0.01 & 0.15 &  0.29$\pm^{0.10}_{0.07}$ & (0.16) \\
Norm.$^{7}$ & 10.08$\pm^{5.77}_{3.59}$& 16.92 & 0.07$\pm^{0.16}_{0.05}$ & (10.07) \\
$\chi^{2}$/d.o.f. & 44.10/26 & 81.60/22 & 13.04/18 & 18.25/19 \\         
\hline\\
DISKBB+PL$^{4}$ & & & &\\
$N_{H}$$^{5}$ & (0.016) & (0.016) & (0.016) & \\
$kT_{in}$$^{6}$ & 0.16$\pm$0.01 & 0.11$\pm$0.01 &  0.29$\pm^{0.12}_{0.99}$ &\\
Norm.$^{7}$ & 11.24$\pm^{7.36}_{4.37}$ & 65.33$\pm^{14.55}_{33.28}$ & 0.07$\pm^{0.23}_{0.06}$ &\\
$\Gamma$ & 0.34$\pm^{0.91}_{2.73}$ & 1.61$\pm^{0.42}_{0.24}$ & 1.45 $^{*}$ & \\
Norm.$^{8}$ & 7.86$\pm^{30.3}_{7.75}$ & 52$\pm^{30}_{24}$ & 1.60 $^{*}$ &\\
 $\chi^{2}$/d.o.f. & 36.07/24 & 29.64/20 & 12.98/16 &\\
%DISKPBB & & & \\
%$N_{H}$ & (0.016) & (0.016) & (0.016)\\
%$kT_{in}$ & 1.78 & 0.26$\pm^{0.24}_{0.06}$ & 0.89\\
%$\rho$ & 0.5 & $>$0.53 & 0.5\\
%Norm. & 8.91$\times$10$^{-5}$ & $<$0.20 & 8.71$\times$10$^{-4}$\\
%$\chi^{2}$/dof & 8.86 & 0.74 & 8.81\\
\hline\hline
\multicolumn{5}{l}{1. absorption column is frozen.} \\
\multicolumn{5}{l}{2. absorption column is free to vary.} \\
\multicolumn{5}{l}{3. multiple blackbody disc (MCD) model.} \\
\multicolumn{5}{l}{4. MCD plus power-law model.} \\
\multicolumn{5}{l}{5. absorption column in unit of 10$^{22}$ atoms cm$^{-2}$.} \\
\multicolumn{5}{l}{6. inner accretion disc temperature in unit of keV.} \\
\multicolumn{5}{l}{7. $\left((R_{in}/km)/(D/10kpc)\right)^{2}cos\theta$, where $R_{in}$ is the inner disk radius,} \\ 
\multicolumn{5}{l}{~~~$D$ is the distance to the source, and $\theta$ is the disk declination angle} \\
\multicolumn{5}{l}{8. photons keV$^{-1}$cm$^{-2}$s$^{-1}$ at 1 keV in unit of 10$^{-7}$.} \\
\multicolumn{5}{l}{* see text in section 3.}
%\multicolumn{5}{l}{8. the confidence level is less than 68$\%$, thus no error is quoted.} \\
\end{tabular} 
\end{table}
 
\subsection{X-ray colour}

\begin{figure}
\begin{center}
\includegraphics[width=90mm]{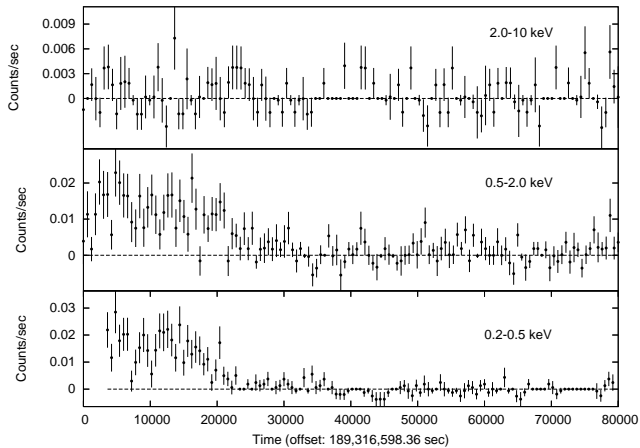}
\caption{The XMM-Newton lightcurves in three colours:0.2-0.5, 0.5-2.0, and 2.0-10.0 keV. The integration time is 600 second and  dash lines in each panels emphasise the zero counts/s level.}
\label{colours}
\end{center}
\end{figure}

X-ray colour is a useful tool to reveal spectral variability in X-ray
binaries \citep[see e.g.][]{van-der-Klis:2006dp}. We created X-ray lightcurves of
the EPIC-pn camera in three energy ranges: 0.2-0.5 keV, 0.5-2.0 keV,
and 2.0-10.0 keV (see Figure \ref{colours}). The X-ray flux decline is
clearest in the soft energy (i.e below 2 keV), as already indicated by
the spectral differences between the bright and faint epochs. In the high
energy band (2.0-10.0 keV), the source was scarcely detectable and has no obvious variation throughout the observation.
 
\section{Discussion}
\label{discuss fitting}
A few key results come from the spectral fitting:\\
{\it First}, it suggests that bright epoch XMM-Newton and Chandra spectra can generally be described by a conventional MCD model with similar characteristic temperature $kT_{in}$. The X-ray state of the object during the Chandra observation is consistent with the bright epoch XMM-Newton.\\ 
{\it Second}, the slim disc-like $\rho$-free model fails to be an alternative interpretation of the spectra in both observations, especially in the brighter epochs. \\
{\it Third}, the fitting result and X-ray colour do not support the scenario that the X-ray decline is caused by an X-ray state change. X-ray transients, many of which are black hole X-ray binaries, exhibit distinct X-ray spectral behaviour at different X-ray flux. In the high state, the soft, thermal component dominates the spectrum, while in the low state, this component reduces to less than 20$\%$, and the hard component becomes dominant $>$80$\%$ \citep[see][]{Remillard:2006fk}. We did not see any such qualitative changes in the spectrum between the bright and the faint epochs. Additionally, these state transitions are not typically seen to take place in a few hours in stellar mass black holes, but rather over days or weeks. Because these transitions typically happen at 2\% of the Eddington luminosity \citep{Maccarone:2003bh}, the black hole mass would have to be about $1000 M_\odot$ for the observed spectral variability to be soft-to-hard state transition, and the increase in mass by a factor of 100 would be expected to lead to an increase in characteristic variability timescale by a factor of 100. For these reasons, we can be confident that the observed spectral variability is due to a subtler cause than spectral state transitions. \\
{\it Fourth}, We consider the possibility that the X-ray decline is an actual eclipse, and the faint epoch XMM-Newton data represent the residual emission coming from other sources within the globular cluster. This spectrum is soft ($kT\sim$ 0.29 keV) and its X-ray luminosity in the 0.2-10 keV band is $\lesssim$10$^{39}$ erg s$^{-1}$. If the residual emission is from a collection of LMXBs, then these LMXBs will be much softer than the typical galactic LMXB. On the other hand, it is not bright enough to be classified as an ULX \citep[$>$10$^{39}$ erg s$^{-1}$, see][]{Fabbiano:2006kx}. No currently known stellar-type X-ray sources has such exotic spectral characteristics. Furthermore, the residual emission of eclipsing X-ray binaries, such as Her X-1 and extra-galactic X-ray binary M33 X-7 \citep{1988MNRAS.231...69D, 2006ApJ...646..420P}, is at most $\sim$10\% of the uneclipsed flux, and can be explained by reprocessing of primary photons from the compact X-ray source in an extended accretion disc corona or by scattering in the companion atmosphere/stellar wind. On the other hand, at flux minimum, the black hole candidate is still luminous, and appears to be only absorbed, rather than to be scattered light. As a result we suggest that the eclipse scenario is not a suitable explanation.
 
\subsection{Nature of the decline}
An increasing absorption column is thus a more consistent and simpler
explanation of the X-ray decline so far. A few key possibilities for
the origin of the excess absorption are a stellar wind from the donor
star (although this should reveal sinusoidal modulation of $N_H$
around the entire orbit, contrary to observations), a grazing eclipse
by the donor star, a disc wind from the accretion disc around the
black hole, or a grazing eclipse by a precessing, warped accretion
disc. A strong diagnostic of the possible cause is the timescale of the X-ray decline. To obtain the timescale $\Delta t$, the lightcurve in the 0.2-2.0 keV band was fitted by three straight line segments to find out two vertices: {\it first} ($t_{1st}$) and {\it second} ($t_{2nd}$) contacts (here we just borrow the terminology of eclipse), thus $\Delta t=t_{2nd}-t_{1st} \sim 12600\pm1600$ s. 

First we test if this is an occultation by a Roche lobe filling
secondary star by calculating the possible orbital period $P$
depending on $\Delta t$ and the size of the X-ray emission region
$D=2R_{in}\approx10000$ km derived from the MCD model. Assuming a circular
orbit as the simplest case, the ingress time of $\Delta t$ that the
secondary star crossing over the linear dimension of the X-ray source
$D$ is proportional to the binary separation $a$ and the orbital
period $P$,

\begin{equation}
\label{rotation}
\frac{2\pi a}{P}=\frac{D}{\Delta t}
\end{equation} then combining with the Kepler's third law, we derive the equation for the orbital period: \\

\begin{equation}
P=2\pi GM\left(\frac{\Delta t}{D}\right)^{3} ,
\end{equation} where $M$ is the total mass of the binary system, and $G$ is gravitational constant. \\

For a X-ray binary containing a stellar mass black hole, $M \simeq$10$M_{\odot}$, much of the total mass of the binary can be attributed to the BH. From equation 2 above it is clear that this leads to an unphysically
long orbital period of $\sim10^{9}$ hour. This not only refers to an
unrealistic size for a Roche lobe overflowing binary system \citep{Eggleton:1983ck}, but also
means that the size of semi-major axis of the orbit is at a similar
scale to the tidal radius of typical globular clusters. Increasing the mass of the BH to say an intermediate mass black hole compounds the problem by increasing the inferred orbital period. 

Therefore, and eclipse by the donor star cannot explain the change in
flux.

\subsection{Precessing warped accretion disc?}
From both the observational and theoretical viewpoints, it is possible
that the X-ray source can be quasi-periodically obscured by a tilted
accretion disc precessing around it roughly on a long period
$P_{long}$, such as Her X-1, SS 433, and LMC X-4
\citep[see][]{White:1995ex}. Accretion discs can be tilted by the
tidal force of the companion star \citep[i.e. in Cataclysmic Variables
(CVs),][]{Whitehurst:1988vy}, or warped if the central radiation
source exerts non-axisymmetric radiation pressure on an initially flat
disc, and the warping is driven in the outer part of accretion disc
\citep[i.e. in X-ray binaries][]{Pringle:1996cq}. The precession
period of a warped accretion disc is much longer than the orbital
period of the binary system \citep{Wijers:1999cu}, e.g. the orbital
period of Cyg X-1 is 5.6 days and its precession period is 294 days
\citep{Priedhorsky:1983fk}. It is clear that XMMU 122939.7+075333 generally
fulfils the conditions required for a precessing warped accretion disc
- as one of the brightest known X-ray binaries, its X-ray luminosity
is sufficiently strong to induce warping.  Additionally, given its
long outburst, it is likely to have a long orbital period, and long
period systems are more susceptible to irradiation driven warping than
short period systems.

Next consider whether the precession period that would be
required to produce the observed results are reasonable.  Applying
equation \ref{rotation}, and solving for $P$, we estimate that a $\sim$90
day precession period would be inferred for a warp located about
$10^{11}$ cm from the black hole (a reasonable distance if we assume an orbital period of $\sim10^{0}$ days, also a stellar mass black hole). This is
well in the range of precession periods from the currently known X-ray
binaries \citep[see][]{Ogilvie:2001kk}. While the ratio of precession
period to orbital period is a bit lower than that typically seen from
X-ray binaries, the precession period is expected to be shorter for
higher X-ray luminosities \citep{Maloney:1997qe,Wijers:1999cu}, so
again, the data are, at the very least, consistent with this picture's
predictions.  We note that if the black hole is of intermediate mass,
the characteristic radius will be considerably larger than $10^{11}$
cm, requiring a similarly longer precession period.

A few alternative hypotheses exist. For example, the change in
absorption could come from an outflow generated by super-Eddington
accretion \citep[e.g.][]{Proga:1998fk,Begelman:2006kx}. If the wind
itself is variable, then different column densities could be in front
of the accretor at different times. This would be a possibility in
either the case where the system shows long outbursts due to a highly
evolved donor star, or where the system is truly persistent, due to
accretion from a white dwarf with a sufficiently short orbital period
that the entire accretion disc is ionised. However, it would
require some degree of fine tuning for the spectrum to be consistent
with only foreground absorption during the part of the observation
when the source is bright, while changing by a factor of about 10 when
the source becomes fainter.

\section{Conclusion}
We analyse the X-ray spectrum of the black hole X-ray binary
candidate XMMU 122939.7+075333 from the XMM-Newton and Chandra
observations. The variation in the XMM-Newton observation is dramatic
but the X-ray colour lacks the spectral signature usually seen
in X-ray state change. 

%The residual emission in the XMM-Newton data is soft ($<$ 1 keV) and still luminous ($\sim$90\% of the bright state). This is contrast to the eclipsing extra-galactic X-ray binary M33 X-7. Although, during the eclipse, M33 X-7 shows an ingress timescale that is similar to the candidate, its residual emission is $\sim$4\% of the uneclipsed flux, and can be explained by reprocessing of primary photons from the compact X-ray source in an extended accretion disc corona or by scattering in the companion atmosphere/stellar wind \citep{2006ApJ...646..420P}.

The event can simply be explained by an
increase of an absorption column, as MKZR07 has mentioned earlier. The
spectral properties of the Chandra data are similar to those of the
bright epoch XMM-Newton data, suggesting it is the ``normal''
state for the object. According to the timescale of the variation in
the X-ray luminosity during the XMM-Newton observation, we suggest
that a precessing warped accretion disc is the source to obscure the
central X-ray emission region.

As a warped disc surface can tilt off from the orbital plane for up to
40$^{\circ}$ \citep{Wijers:1999cu,Ogilvie:2001kk}, an observer must be
at a high enough inclination angle ($>$50$^{\circ}$-70$^{\circ}$) to
see the warped accretion disc crossing his line of sign to the central
X-ray source. If the object has a rather stable precessing warped
accretion disc, we should be able to see more similar variation events
by carefully programming further observations to the timescale of the
precession period. The most plausible alternative model is that a
disc wind from the accretion disc is varying. A monitoring campaign
on this source could discriminate between the two possibilities, since
the variations in $N_H$ should follow a regular pattern if they are
due to precession of the accretion disc, and should be aperiodic,
rather than quasi-periodic, if they are due to a variable disc wind.

\section{Acknowledgements}
ICS thanks the University of Southampton for the hospitality while a
part of this work was done. AK and SEZ acknowledge support for this
research from XMM NASA grant no NNG04GF54G and Chandra grant SAO
AR7-8010X. AK was supported by NASA-LTSA grant NAG5-12975. We thank
Robin Barnard, Phil Charles, Guillaume Dubus \& Steinn Sigurdsson for useful discussion. We also want to thank the referee for helpful comments on improving this paper.

\end{document}